\documentclass[onecolumn,aps,prd,preprintnumbers,showpacs,superscriptaddress,nofootinbib,amsmath,amssymb,floats,floatfix,showkeys,notitlepage,longbibliography]{revtex4-1}

\usepackage{comment}
\usepackage{graphicx}
\usepackage{subfigure}
\usepackage{palatino}
\usepackage{sans}
\usepackage{hyperref}
\hypersetup{colorlinks=true,linkcolor=blue,urlcolor=blue,citecolor=blue}
\usepackage[toc,page]{appendix}
\usepackage[normalem]{ulem}
\usepackage{adjustbox}
\usepackage{latexsym}
\usepackage{amsmath}
\usepackage{amssymb}
\usepackage{amsfonts}
\usepackage{times}
\usepackage{dcolumn}
\usepackage{bm}
\usepackage{tikz}
\usepackage{bigints}
\usepackage{array,tabularx,multirow,booktabs}
\usepackage{color,soul} 
\usepackage[tracking=true]{microtype}
\SetTracking{}{500}
\SetTracking{encoding={*}, shape=sc}{40}
\UseRawInputEncoding 
\allowdisplaybreaks
\DeclareMathOperator\sech{sech}

\begin{document} \sloppy

\title{\textbf{Weak Deflection Angle, Hawking Radiation and Greybody Bound of Reissner-Nordstr\"{o}m Black Hole Corrected by Bounce Parameter}}

\author{Wajiha Javed}
\email{wajiha.javed@ue.edu.pk} 
\affiliation{Department of Mathematics, Division of Science and Technology, University of Education, Lahore-54590, Pakistan}

\author{Mehak Atique}
\email{mehakatique1997@gmail.com} 
\affiliation{Department of Mathematics, Division of Science and Technology, University of Education, Lahore-54590, Pakistan}

\author{Reggie C. Pantig}
\email{rcpantig@mapua.edu.ph}
\affiliation{Physics Department, Map\'ua University, 658 Muralla St., Intramuros, Manila 1002, Philippines}

\author{Ali {\"O}vg{\"u}n}
\email{ali.ovgun@emu.edu.tr}
\affiliation{Physics Department, Eastern Mediterranean University, Famagusta, 99628 North
Cyprus via Mersin 10, Turkey.}

\begin{abstract}
In this study, we probe the weak lensing by a Reissner--Nordstr\"{o}m black hole corrected by bounce parameter in plasma and dark matter mediums. For this, the optical geometry and the Gibbons--Werner approach are utilized to obtain the bending angle in the weak field limitations. We examine that the impact of these mediums increases the black hole's bending angle. In addition, we graphically study the deflection angle of light with respect to the impact parameter and examine that the bounce parameter directly affects the angle. Further, we compute the Hawking radiation via a topological method involving two invariants and verify our obtained result with the standard method of calculating the Hawking temperature. In addition, we compute the greybody factor's bound of the black hole. Moreover, we analyze the bound graphically and observe that the bound shows convergent behavior. We also study that our attained results reduce the results of the Reissner--Nordstr\"{o}m and Schwarzschild black holes by reducing the parameters. Finally, we probe how the bounce parameter affected the shadow radius and compared it to the shadow produced if the black hole is immersed in plasma. It is revealed that the rate at which the shadow radius changes with respect to $r$ easily tends to zero under the effect of the bounce parameter, while the plasma merely increases the shadow radius.

\end{abstract}



\date{\today}
\maketitle

\section{Introduction}
General relativity (GR) is the theory of gravity proposed by Einstein in $1916$. In GR, Einstein gave the idea of black holes (BHs) \cite{Einstein:1936}. Acknowledging Newton's corpuscular theory of light, which assumed that photons {are ultra-light} particles, geologist Michell proposed the presence of dark stars. Today, these dark stars are known as BHs. Black holes are fascinating astronomical objects with a gravitational attraction so powerful that nothing can escape them, not {even light}. According to the no-hair theorem, all astrophysical BHs are fully defined by their masses and spins. A BH has two main components: the singularity and the event horizon. The question which attains the most attention in GR is about the inner structure of a BH. However, because of the presence of the spacetime singularity, where the curvature deviates continuously, and GR breaks down, this question cannot be answered easily. The singularity theorems, presented by Penrose and Hawking~\cite{Penrose:1965}, state that gravitational collapse with physically valid circumstances always results in the formation of a singularity. Even in some scenarios (such as {a} cosmological constant in the spacetime area), the singularity theorem's assumptions may not be applied. Black holes having regular centers or, in other words, having no singularity are called regular BHs or non-singular BHs. The first regular BH with horizons and no core singularity~\cite{Ansoldi:2008}  was proposed by Bardeen~\cite{Bardeen:1968}. Near the origin, the Bardeen~BH~behaves like a de Sitter spacetime, however, for  $r \rightarrow \infty$, it acts like a Schwarzschild BH \cite{Ansoldi:2008}. Later, Ayon-Beato and Garcia~\cite{Ayon:1999} showed that Bardeen's model is an accurate solution of GR connected to non-linear electrodynamics. There has been significant progress in the study and application of regular BHs \cite{Dymnikova:1992,Lemos:2011,Kumar_2020}, as well as regular rotating {black holes} \cite{Kumar_2020,Eichhorn:2021}. Most of these solutions are based on Bardeen's concept, which uses non-linear electrodynamics as a source.

According to Hawking \cite{HAWKING:1975np}, a BH can emit heat radiation by taking into account the quantum consequences, and Hawking radiation is a term used to describe this type of heat radiation. The production and annihilation of particles are theoretically feasible in the context of quantum field theory. When pair creation occurs near the BH's horizon, one of the particles from the pair falls back to the BH while the other particles depart from the BH's horizon. The particles that exit are observed by an outside observer as Hawking radiation \cite{Hassanabadi:2021kyv,Hassanabadi:2021kuc,Chen:2022dap}. According to GR, the spacetime bend by a BH acts as a gravitational potential inside which the particles move. Some Hawking radiation is returned to the BH while the remainder passes through the potential at infinity. In this aspect, the transmission probability is known as the greybody factor. Several methods for obtaining Hawking radiation have been proposed. Using a topological method, Zhang, Wei, and Liu \cite{Peng:2020xnr} investigated the Hawking temperature of the BTZ BH.  {\"O}vg{\"u}n et al. \cite{Ali:2020} computed the Hawking temperature for BHs by applying the topological strategy. Kruglov \cite{Kruglov:2018} explored the Hawking temperature of a magnetically charged BH through surface gravity and horizon in the context of non-linear electrodynamics.
The greybody factor can be calculated in a variety of ways. The matching approach can be used to derive an estimated greybody factor \cite{Fernando:2005,Kim:2008,Jorge:2008}. The WKB approximation may be used to calculate the greybody factor if the gravitational potential is high enough \cite{Parikh:2000,Fleming:2005}. The greybody factor may also be calculated using the rigorous bound, which is an alternative to approximation. The bound can be used to describe a BH qualitatively. Visser \cite{Visser:1999} inspected some extremely wide reflected and transmitted coefficient constraints for one-dimensional potential scattering. Boonserm and Visser \cite{Boonserm:2008} calculated the greybody factor's bound for Schwarzschild BHs by examining the Regge-Wheeler equation for wave phase angular momentum and arbitrary particle spin. Javed, Hussain, and  {\"O}vg{\"u}n \cite{Javed:2022} worked out the boundaries of the Kazakov SolodukhinBH's greybody factor.

The gravitational lensing (GL) effect states that a light beam would be distorted while passing by a huge object, {which is} one of GR's most important predictions. For determining the mass of galaxies and clusters \cite{Henk:2013,Brouwer:2018}, as well as discovering dark energy and dark matter (DM) \cite{Vanderveld:2012}, GL has become one of the most powerful instruments in astronomy and cosmology. Since the first measurements of the Sun's gravitational bending of light, the lens equation has been used to examine the GL effects for BHs, wormholes, cosmic strings, and other objects. Strong GL and weak GL are the types of GL. Strong GL is a GL effect that is intense enough to generate many pictures, such as arcs or Einstein's rings. In this type of GL, geometry is favorable and bending is rather large, whereas the weak GL is a GL impact that is not intense to create multiple pictures, and the geometry is less suitable. Since the 19th century, various studies on the GL have been done not only for the BHs but also for the wormholes, cosmic strings, global monopoles, {and} neutron stars \cite{Keeton:1998,Bhadra:2003,Whisker:2005,Chen:2009,Nandi:2006,Eiroa:2002,
Kumaran:2019qqp,sym14102054}.

Gibbons and Werner (GW) presented a technique for calculating {the} deflection angle in $2008$. The Gauss-Bonnet theorem (GBT) and the optical geometry of the BH's spacetime, where the source and~viewer~are in asymptotic areas, were used to develop their technique. Werner \cite{MC:2012} soon expanded this approach to stationary spacetimes. In GBT, one can utilize the domain $\mathcal{G_{S}}$ confined by the light ray as well as a circular boundary curve $\mathcal{C_{S}}$ placed at the lens's center~where the light ray intersects at the source and~receiver. The source and receiver are considered at the same coordinate distance $\mathcal{S}$ from the lens. The GBT in optical metric and by using weak field approximation can be expressed as follows 
\begin{equation}
 \int\int_{\mathcal{G}_{S}}\mathcal{\tilde{K}} dS+\oint_{\partial\mathcal{G}_{S}}kdt
 +\sum_{i}\epsilon_{i}=2\pi\mathcal{X}(\mathcal{G}_{S}),\label{w1}
\end{equation}
wherein $\mathcal{\tilde{K}}$ indicates the Gaussian optical curvature, $k$ stands for geodesic curvature, $dS$ is a surface element of optical geometry, and $\mathcal{G_{S}}$ is a region accommodates the light rays source, the referential of the observer and the lens's center. {At $i$th vertex, $\epsilon_i$ represents the exterior angles.} Just for simplicity, we suppose that as long as radial distance $S\rightarrow\infty$, the sum of the external angles $\theta_{i}$ becomes $\pi$ for the observer. The asymptotic deflection angle $\tilde{\alpha}$ can be calculated as:
\begin{equation}
\tilde{\alpha}=-\int_{0}^{\pi} \int_{\frac{b}{\sin(\phi)}}^{\infty} \mathcal{\tilde{K}} dS,\label{w2}\\
\end{equation}
where $b$ represents the impact parameter. The integral of GBT may be solved in an infinite region confined by a ray of light. Instead of utilizing the asymptotic receiver and source, Ishihara et al. \cite{Suzuki:2016,Asahi:2017} modified this approach for finite distances. The finite-distances approach was then applied to the axisymmetric spacetimes by Ono et al. \cite{Asada:20117}. In {the} plasma medium, the GBT was used by Crisnejo, and Gallo \cite{Crisnejo:2018} to determine the gravitational bending of light. Using massive particles and the Jacobi--Maupertuis Randers--Finsler metric inside GBT,~Li et al. \cite{Zonghai:2020,Guodong:2020} investigated the finite-distance impacts on the weak bending~angle.

Beginning from the first startling discoveries by Oort \cite{Oort:1932} of missing matter in the Galactic disk, which modern observations have not confirmed, and by Zwicky, \cite{Zwicky:1937} the discovery of missing matter in the Coma cluster, much later understood to be ``DM". Dark matter comprises particles that do not absorb, reflect, or emit light, making it impossible to detect them using electromagnetic radiation. Only gravitational interactions can detect DM, and we know that DM is non-baryonic, non-relativistic, and possesses weak non-gravitational interactions. Weakly interacting massive particles (WIMPs), super-WIMPs, axions, and sterile neutrinos are the four candidates of DM \cite{Feng:2010}. Dark matter constitutes about $85\%$ of the total mass of the Universe \cite{Jarosik:2011} and is used to explain the strange behavior of stars and galaxy dynamics. In DM medium, Pantig and  {\"O}vg{\"u}n \cite{Ovgun:2019, Pantig:2022toh,Pantig:2020odu,Pantig:2022sjb,Pantig:2022whj} studied the weak GL by wormholes and BHs.

The light that passes close to the BH is refracted by the gravitational field, creating the BH's shadow. The BH's shadow is a dark area frequently surrounded by a luminous ring. The BH's mass and angular momentum determine its size and form. Many scientists have attempted to predict how the observable appearance of a BH surrounded by bright material would seem before the spectacular finding of the BH's shadow produced by Event Horizon Telescope Collaboration \cite {EventHorizonTelescope:2019xnr,EventHorizonTelescope:2022xnr}. For instance, Bardeen et al. \cite{1972ApJ...178..347B} examined the shadows of Kerr BH's, while Synge \cite{Synge:1966okc} studied the shadows of Schwarzschild spacetime. The bright accretion disc surrounding the BH was manually drawn by Luminet \cite{Luminet:1979nyg}. In addition,  due to the transparent emissions close to the BH, it is predicted that a BH would display its shadow, which is brought on by gravitational light deflection and photons captured at its event horizon. The photon ring, a geometric characteristic {of spacetime}, determines the shadow radius \cite{Narayan_2019}. To our knowledge, few studies {about RN} black hole with bounce parameter have been conducted. For instance, \cite{Guo:2021wid} has considered the photon rings and shadows. In this study, we ought to analyze such a metric under the influence of plasma and dark matter. With the shadow cast, it can also determine imprints of spacetime, and several studies were also conducted about using the black hole for dark matter detection~\cite{Pantig:2022whj,Pantig:2021zqe, Pantig:2022sjb,Konoplya:2022hbl,Konoplya:2019sns,Xu2018,Xu2021a,Pantig:2020uhp,universe8110599,Javed:2022fsn,Jusufi:2020cpn,Nampalliwar:2021tyz}.

This paper aims to study the weak GL of black bounce Reissner--Nordstr\"{o}m spacetime utilizing optical geometry and GBT in plasma and DM mediums. Moreover, it would be interesting to calculate the Hawking temperature and greybody bound of the BH. We will also study the graphical behavior of {the deflection angle} and {greybody bound}. We will focus on how the bounce parameter (introduced in Reissner--Nordstr\"{o}m BH) affects the bending angle, Hawking temperature, and the bound.

The layout of our paper is given as follows. Section \ref{sec2} is based on the discussion about the black bounce Reissner--Nordstr\"{o}m spacetime. In Section \ref{sec3}, we obtain the optical metric from a four-dimensional spherically symmetric metric and then compute the deflection angle of the BH {with plasma medium} by using GBT, and analyze its graphical {behavior}. The computation of the bending angle in the case of DM medium is given in Section \ref{sec5}. Section \ref{sec6}, is devoted to investigating the Hawking temperature of black bounce Reissner--Nordstr\"{o}m BH via GBT. The computation of the greybody factor's bound of black bounce Reissner--Nordstr\"{o}m BH, and the graphical behavior of the bound is addressed in Section \ref{sec7}. Finally, Section \ref{sec8} discusses the shadow behavior. The purpose of Section \ref{conclu} is to sum up the findings of this research and propose a research direction. Throughout the paper, we used natural units $G = c = 1$ and metric signature $(-,+,+,+)$.

\section{Black Bounce Reissner-Nordstr\"{o}m Spacetime} \label{sec2}
One of the most significant {problems} is the prediction of the spacetime singularity within a BH or at the start of the universe, which implies that the GR theory fails there. To address the singularity issue,  Bardeen \cite{Bardeen:1968}  was the first who put up the idea of regular BHs, which has continuously attracted scientific interest. It is convenient to consider the regular BH due to the problematic nature of spacetime’s singularities. Regular BHs are solutions of the gravity equations that have an event horizon but no singularities {in spacetime. Based on the bounce} and quantum corrections, a wide range of regular black hole solutions have been attained \cite{Thomas,Hyat,Franzin}. The black bounce spacetime smoothly interpolates among ordinary Schwarzschild BH and Morris--Thorne traversable wormhole \cite{22}. It is noteworthy to remark that throughout the geometry is regular { and one can {have} a distinct type of ``regular BH'', where the ``origin'' $r = 0$ {can either} be spacelike, null, or timelike, as long as the parameter $a \ne 0$.} Additionally, it was demonstrated that the spacetime metric may be utilized to characterize {several} interesting physical circumstances, such as a developing black-bounce, a wormhole to black-bounce transfer, and the opposite black-bounce to wormhole transition. In Reissner--Nordstr\"{o}m BH, a regularizing~process was recently proposed \cite{Franzin}, which does not produce a standard regular BH \cite{Hayward:206}, such as the Bardeen or Hayward BHs, rather, it produces a charged regular BH called a black bounce Reissner--Nordstr\"{o}m or charged black bounce. In a static spherically symmetric spacetime, the line-element for the black bounce Reissner--Nordstr\"{o}m BH can be described as \cite{Guo:2022hjp}
\begin{equation}
ds^2=-f(r)dt^2+ \frac{dr^2}{f(r)}+h(r)^2(d\theta^2+\sin^2\theta d\phi^2),\label{M1}
\end{equation}
where the metric function $f(r)$ and $h(r)^2$ are defined as
\begin{equation} \label{emetfunc}
f(r)=1-\frac{2m}{\sqrt {r^{2}+a^{2}}}+\frac{Q^{2}}{r^{2}+a^{2}}~~~~\text{and}~~~~h(r)^2=r^{2}+a^{2}.\nonumber
\end{equation}

In the metric function, 
 $m$ stands for the mass of the BH, $Q$ indicates the charge, and $a$ represents the bounce parameter of black bounce Reissner--Nordstr\"{o}m. Several characteristics of the black bounce family have been thoroughly investigated \cite{Lobo:2021,Simpson_2019}. Some properties are that the curvature singularities are absent from the black bounce family on a global scale and satisfy all observable weak field tests. Based on the values of charge $Q$ and bounce parameter $a$, one can easily interpolate the Reissner--Nordstr\"{o}m and Schwarzschild BHs. By taking charge $Q \ne 0$ and bounce parameter $a=0$, one can obtain the Reissner--Nordstr\"{o}m BH. If we consider the charge $Q=0$ and bounce parameter $a=0$, then we can obtain the Schwarzschild BH. The event horizon of the Reissner--Nordstr\"{o}m BH corrected by bounce parameter can be computed by taking $f(r)=0$.
\begin{equation}
r_{h}=\sqrt{(m+\sqrt{m^{2}-Q^{2}})^{2}-a^{2}}.\nonumber
\end{equation}

One can observe 
that a coordinate speed of light may be described in terms of radial null curves $(ds^{2} = d\theta = d\phi = 0)$, since the radial coordinate $r \in (-\infty, +\infty)$ \cite{Guo:2022hjp}:
\begin{equation}
    C(r)=\left\lvert \frac{dr}{dt} \right\lvert=f(r)=1-\frac{2m}{\sqrt {r^{2}+a^{2}}}+\frac{Q^{2}}{r^{2}+a^{2}}. 
\end{equation} 

Thus in this spacetime, a sphere's area at $r$ (radial coordinate) has the following form $A(r)=4\pi h(r)^2$. The wormhole throat is where the area is minimized, and by observing the state, one may determine where the throat is {$A'(r_{o})=0$}, where the throat’s location is represented by $r_{0}$. The wormhole throat's radius is thus given by $h_{0}= h(r_{0})$.  We now divide this geometry into three categories \cite{Guo:2022hjp}:

1. The outer and inner horizon exist at $r_{h}$ for $a < (m+\sqrt{m^{2}-Q^{2}})$ and $\mid Q \mid < m$. In this instance, $\exists$ $r_{h}$ $\in$ $R^{*}$ and $c(r_{h})=0$. Since light has a zero coordinate speed, it cannot escape the horizon. This {geometry indicates} a charged regular black hole with usual outer and inner horizons.

2. One can obtain one extremal horizon $r_{h}=0$,  when $a=(m+\sqrt{m^{2}-Q^{2}})$ and $\mid Q \mid < m$, and we know $\exists$ $r_{h}=0$ and $c(r_{h})=0$. For this case, the geometry represents the extremal charged regular BH, {which is the one-way} charged traversable wormhole with single extremal null throat at $r_{h}=0$.

3. For the case when, $a > (m+\sqrt{m^{2}-Q^{2}})$ and whether {$\mid Q \mid < m$ or  $\mid Q \mid > m$} , there is no horizons. So, we have $\forall$ radial coordinate $r \in (- \infty, +\infty)$ and $c(r) \ne 0$. This case represents a two-way charged traversable wormhole and the light can travel throughout the domain.

\section{Plasma Influenced Deflection Angle} \label{sec3}
Guo and Miao \cite{Guo:2022hjp} have calculated the deflection angle by Reissner--Nordstr\"{o}m BH corrected by bounce parameter in non-plasma medium utilizing GBT. Now, in this section, we see how the presence of a plasma affects the bending of light by black bounce Reissner--Nordstr\"{o}m BH specified by charge $Q$ and bounce parameter $a$. In the scenario of plasma medium, the refractive index for the black bounce Reissner--Nordstr\"{o}m is described as \cite{Crisnejo:2018}.
\begin{equation}
n(r)=\sqrt{{1-\delta\left(1-\frac{2m}{\sqrt {r^{2}+a^{2}}}+\frac{Q^{2}}{r^{2}+a^{2}}\right)}},\label{e4}
\end{equation}
where, $\delta=\frac{\omega^{2}_{e}}{\omega^{2}_{\infty}}$ and the plasma parameters $\omega_{e}$ and $\omega_{\infty}$ represents electron plasma frequency and photon frequency as marked by the static investigator at infinity. For static spherically symmetric metric in Equation $\eqref{M1}$ assuming that the source of light and viewer are on the equatorial region $(\theta = \frac{\pi}{2})$. Because we are working with null geodesics, we use $(ds^{2}=0)$ to find the appropriate optical metric.
\begin{equation}
{optical metric=n^2}dt^{2}=g_{pq}dx^{p}dx^{q}=n^{2}(r)\left[\frac{1}{(f(r))^{2}}dr^{2}+\frac{h(r)^2}{f(r)}d\phi^{2}\right],\label{R3}
\end{equation}
where, $p$,$q$ $\in$ $\{1,2\}$. In order to determine the optical Gaussian curvature $\mathcal{\tilde{K}}$ from the optical metric Equation $\eqref{R3}$, we use the following expression
\begin{equation}
\mathcal{\tilde{K}}=\frac{R}{2},\label{R44}
\end{equation}
where $R$ is the Ricci scalar {calculated using the optical metric}. Utilizing Equation $\eqref{R44}$ the Gaussian optical curvature of the black bounce Reissner--Nordstr\"{o}m BH in plasma medium is calculated as
\vspace{6pt}
\begin{eqnarray}
\mathcal{\tilde{K}} &\simeq& -\frac{2m}{r^{3}}+\frac{3Q^{2}}{r^{4}}-\frac{6m Q^{2}}{r^{5}}+\frac{5Q^{2}\omega^{2}_{e}}{r^{4}\omega^{2}_{\infty}}-\frac{3m \omega^{2}_{e}}{r^{3}\omega^{2}_{\infty}}\nonumber
\\&-&\frac{26m Q^{2}\omega^{2}_{e}}{r^{5}\omega^{2}_{\infty}}-\frac{12a^{2} Q^{2}}{r^{6}}-\frac{a^{2}}{r^{4}}+\frac{28a^{2} m Q^{2}}{r^{7}}+\frac{10a^{2}m}{r^{5}}-\frac{20 a^{2} Q^{2}\omega^{2}_{e}}{r^{6}\omega^{2}_{\infty}}\nonumber
\\&-& \frac{a^{2}\omega^{2}_{e}}{r^{4}\omega^{2}_{\infty}} +\frac{115 a^{2} m Q^{2}\omega^{2}_{e}}{r^{7}\omega^{2}_{\infty}} +\frac{31 a^{2} m \omega^{2}_{e}}{2r^{5}\omega^{2}_{\infty}}+\mathcal{O}(m^{2},a^{4},Q^{4}).\label{R4}
\end{eqnarray}

The obtained value of Gaussian optical curvature in Equation $\eqref{R4}$ will be used later to compute the bending angle. Now, to acquire the deflection angle of the black bounce Reissner--Nordstr\"{o}m BH in a plasma medium, we {make use} of the GBT, which is defined as follows \cite{Werner:2008}
\begin{equation}
 \int\int_{\mathcal{G}_{S}}\mathcal{\tilde{K}} dS+\oint_{\partial\mathcal{G}_{S}}kdt
 +\sum_{i}\epsilon_{i}=2\pi\mathcal{X}(\mathcal{G}_{S}).\label{R5}
\end{equation}

As in above equation,
$k$ describes the geodesic curvature which is defined as $k=g(\nabla_{\dot{\gamma}}\dot{\gamma},\ddot{\gamma})$ wherein $g({\dot{\gamma}},\dot{\gamma})=1$, $\ddot{\gamma}$ illustrates the unit acceleration vector. At $i$th vertex, $\epsilon_{i}$ represents the external angle. As $S \rightarrow \infty$, the jump angles becomes $\frac{\pi}{2}$ so that we obtain  $\theta_{o}+\theta_{S}\rightarrow\pi$. Since $\mathcal{G}_{S}$ is a non-singular region, the  Euler characteristic $\mathcal{X}(\mathcal{G}_{S})$ is equal to $1$, and the following result is obtained
\begin{equation}
 \int\int_{\mathcal{G}_{S}}\mathcal{\tilde{K}} dS+\oint_{\partial\mathcal{G}_{S}}kdt
 +\epsilon_{i}=2\pi\mathcal{X}(\mathcal{G}_{S}),\label{R6}
\end{equation}
where $\epsilon_{i}=\pi$ denotes the total angle of jumps, and since $S \rightarrow 0$, the effective element is acquired as
\begin{equation}
 k({E}_{S})=\mid\nabla_{\dot{E}_{S}}\dot{E}_{S}\mid.\label{R7}
\end{equation}

The geodesic curvature's 
radial component is expressed as follows \cite{Werner:2008}:
\begin{equation}
(\nabla_{\dot{E}_{S}}\dot{E}_{S})^{r}=\dot{E}^{\phi}_{S}
\partial_{\phi}\dot{E}^{r}_{S}+\Gamma^{r}_{\phi\phi}(\dot{E}^{\phi}_{S})^{2}.\label{R8}
\end{equation}

For very large $S$, we obtain
\begin{equation}
(\nabla_{\dot{E}^{r}_{S}}\dot{E}^{r}_{S})^{r}\rightarrow\frac{1}{S}.\label{R9}
\end{equation}

It is asserted that the geodesic
 curvature is independent of topological defects, implying that $k({E}_{S}) \rightarrow \frac{1}{S}$. Utilizing the optical metric given in Equation $\eqref{R3}$, one can write $dt = S d\phi$ and have $k({E}_{S})dt=d\phi$. Now, using all {the above-obtained} results and the straight line approximation  $r=\frac{b}{sin\phi}$. The bending angle $\tilde{\alpha}$ can be calculated by using the formula below:
\begin{equation}
\tilde{\alpha}=-\int^{\pi}_{0}\int^{\infty}_{b/\sin\phi}\mathcal{\tilde {K}}dS,\label{R10}
\end{equation}
where $dS=\sqrt{det{g}}drd\phi$. Using Equation $\eqref{R10}$ and the value of optical Gaussian curvature Equation $\eqref{R4}$, the bending angle $\tilde{\alpha}$ of the black bounce Reissner--Nordstr\"{o}m in plasma medium up to the leading order terms is calculated as
\begin{eqnarray}
\tilde{\alpha} &\simeq& \frac{4m}{b}+\frac{2m \omega^{2}_{e}}{b \omega^{2}_{\infty}}-\frac{8m Q^{2}}{3b^{3}}-\frac{3 \pi Q^{2}}{4b^{2}}+\frac{2m Q^{2} \omega^{2}_{e}}{b^{3}\omega^{2}_{\infty}} \nonumber
\\&-&\frac{\pi Q^{2} \omega^{2}_{e}}{2b^{2} \omega^{2}_{\infty}}-\frac{8 a^{2} m}{3b^{3}}+\frac{a^{2} \pi}{4 b^{2}}+\frac{64 a^{2} m Q^{2}}{15b^{5}}+\frac{27a^{2} \pi Q^{2}}{32b^{4}}\nonumber
\\&-&\frac{4a^{2} m \omega^{2}_{e}}{3b^{3}\omega^{2}_{\infty}}-\frac{16a^{2} m Q^{2} \omega^{2}_{e}}{5b^{5}\omega^{2}_{\infty}}+\frac{9a^{2} \pi Q^{2}\omega^{2}_{e}}{16b^{4}\omega^{2}_{\infty}}+\mathcal{O}(m^{2},a^{4},Q^{4}).\label{R11}
\end{eqnarray}

The above deflection angle depends 
on  the mass $m$, charge $Q$, bounce parameter $a$, impact parameter $b$ and the plasma parameters i.e., $\omega_{e}$ and $\omega_{\infty}$. The terms without the bounce parameter $a$ and which contain charge are due to the charged nature of the BH. The remaining terms are due to the corrections with the bounce parameter $a$. For $a=0$, one can find the bending angle $\tilde{\alpha}$ of a Reissner--Nordstr\"{o}m BH in {a plasma} medium. By neglecting the charge and bounce parameter, the obtained angle $\tilde{\alpha}$ reduces to the bending angle of the Schwarzschild BH.~We also observe that the effect of the plasma increases the deflection angle. The bending angle is inversely proportional to the photon frequency, so the bending angle increase by lowering the photon frequency and assuming the electron frequency is fixed. Moreover, one can attain the bending angle in {the case} of non-plasma medium \cite{Guo:2022hjp} if we take $\omega_{e}=0$ or $(\delta\rightarrow 0)$ in the derived deflection angle in Equation $\eqref{R11}$. We also observe that the obtained deflection angle Equation $\eqref{R11}$ is directly proportional to the mass $m$, charge $Q$, bounce parameter $a$, and inversely proportional to the impact parameter $b$.

\textbf{Graphical Behaviour:} 
Now we look into the graphical behavior of the black bounce Reissner--Nordstr\"{o}m BH's deflection angle $\tilde{\alpha}$ {relative to} the impact parameter $b$, for the fixed value of mass $m$ and charge $Q$, while varying bounce parameter $a$ and plasma term.

\begin{figure}[h]
    \centering
\includegraphics[width=0.75\textwidth]{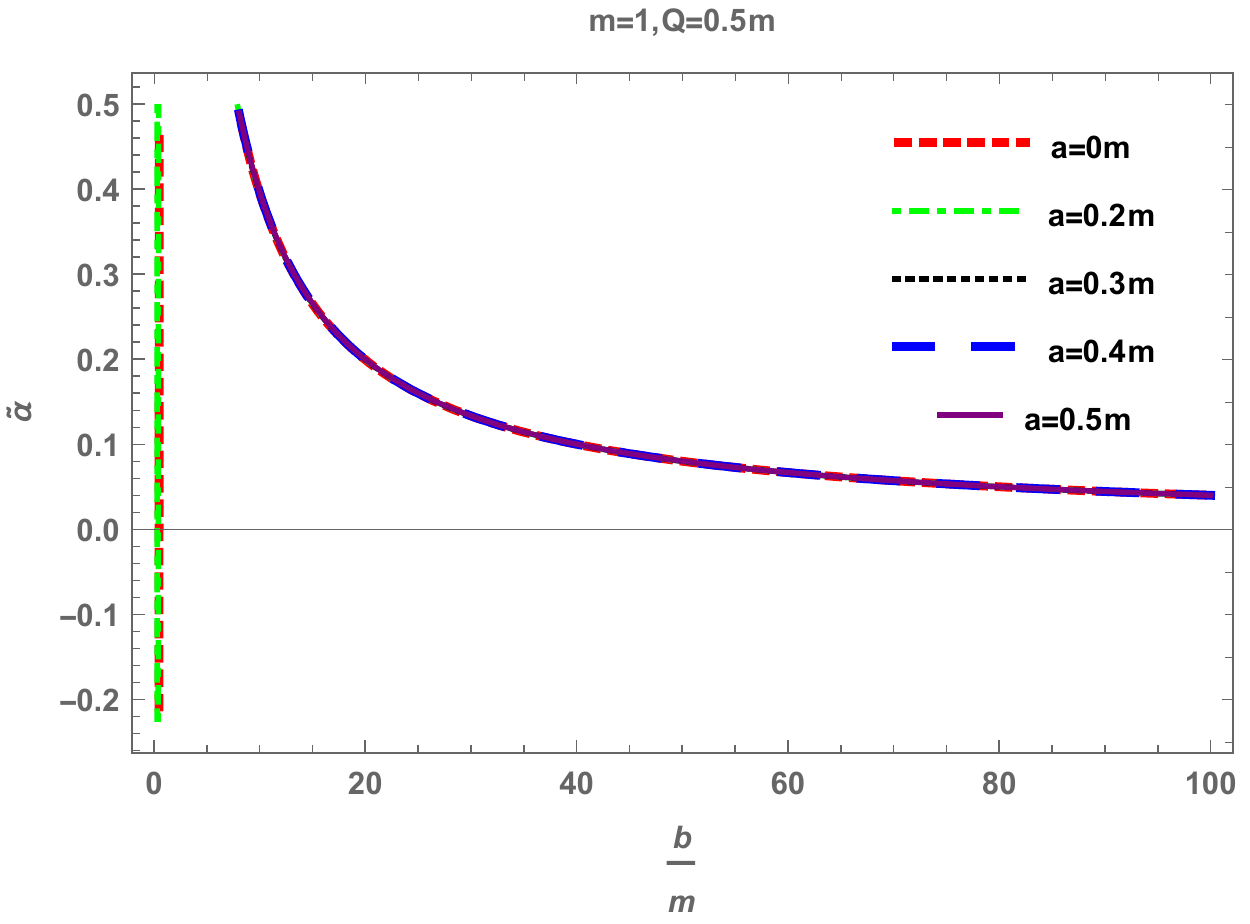}
    \caption{Bending angle's variation $\tilde{\alpha}$ as a function of impact parameter $b$.}
    \label{fig1}
\end{figure}

For fixed values of mass $m$ and charge $Q$, $\frac{\omega_{e}}{\omega_{\infty}}=0.1$ and varying the values of bounce parameter $a$, Figure \ref{fig1} depicts the graph of deflection~angle $\tilde{\alpha}$ vs~impact parameter $b$. {For $a \geq 0$, we investigate that at the small values of impact parameter $b$, one can obtain the maximum value of the bending angle $\tilde{\alpha}$. A the value of $b$ increases, the bending angle $\tilde{\alpha}$ exponentially decreases and {approaches zero}. It is observed that for the small values of $b$, one can obtain the positive angle (deflection in the upward direction).  Further, we examine that the bending angle $\tilde{\alpha}$ shows the inverse {relationship} with the impact parameter $b$.} Moreover, physically the bending angle $\tilde{\alpha}$ represents the stable behavior.

\begin{figure}[h]
    \centering
    \includegraphics[width=0.75\textwidth]{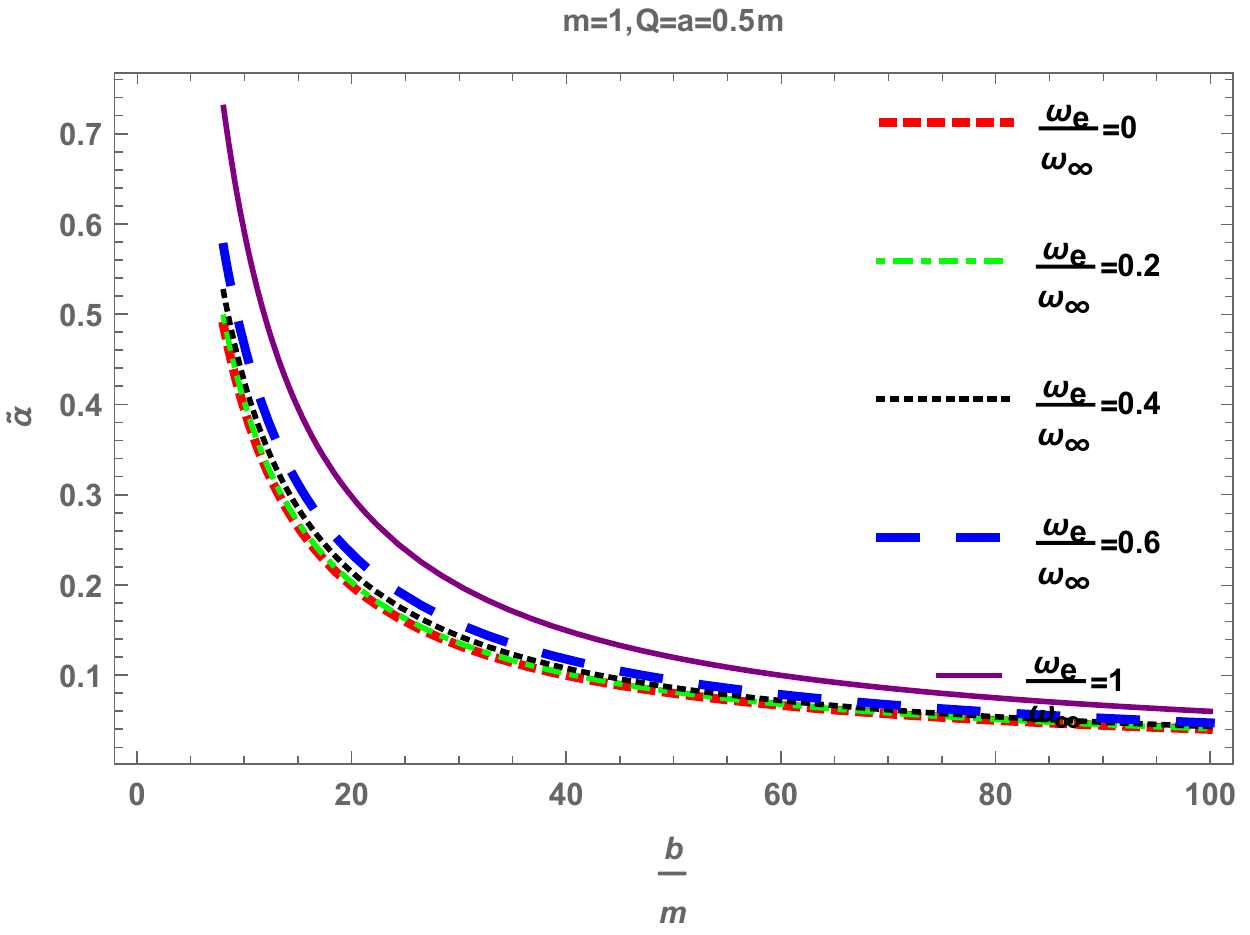}
    \caption{Bending angle's variation $\tilde{\alpha}$ as a function of impact parameter $b$, for $Q=a=0.5m$.}
    \label{fig2}
\end{figure}

Figure \ref{fig2} depicts the behaviour of deflection angle $\tilde{\alpha}$ with respect to the impact parameter $b$  for $Q=a=0.5$ and $0 \leq \frac{\omega_{e}}{\omega_{\infty}} \leq 1$. {We examine that the deflection angle $\tilde{\alpha}$ decreases exponentially and almost approaches {to zero} as the value of impact parameter $b$ goes to infinity. Furthermore, in this case, bending angle $\tilde{\alpha}$ shows the inverse relation with the impact parameter $b$.}

\begin{figure}[h]
    \centering
    \includegraphics[width=0.75\textwidth]{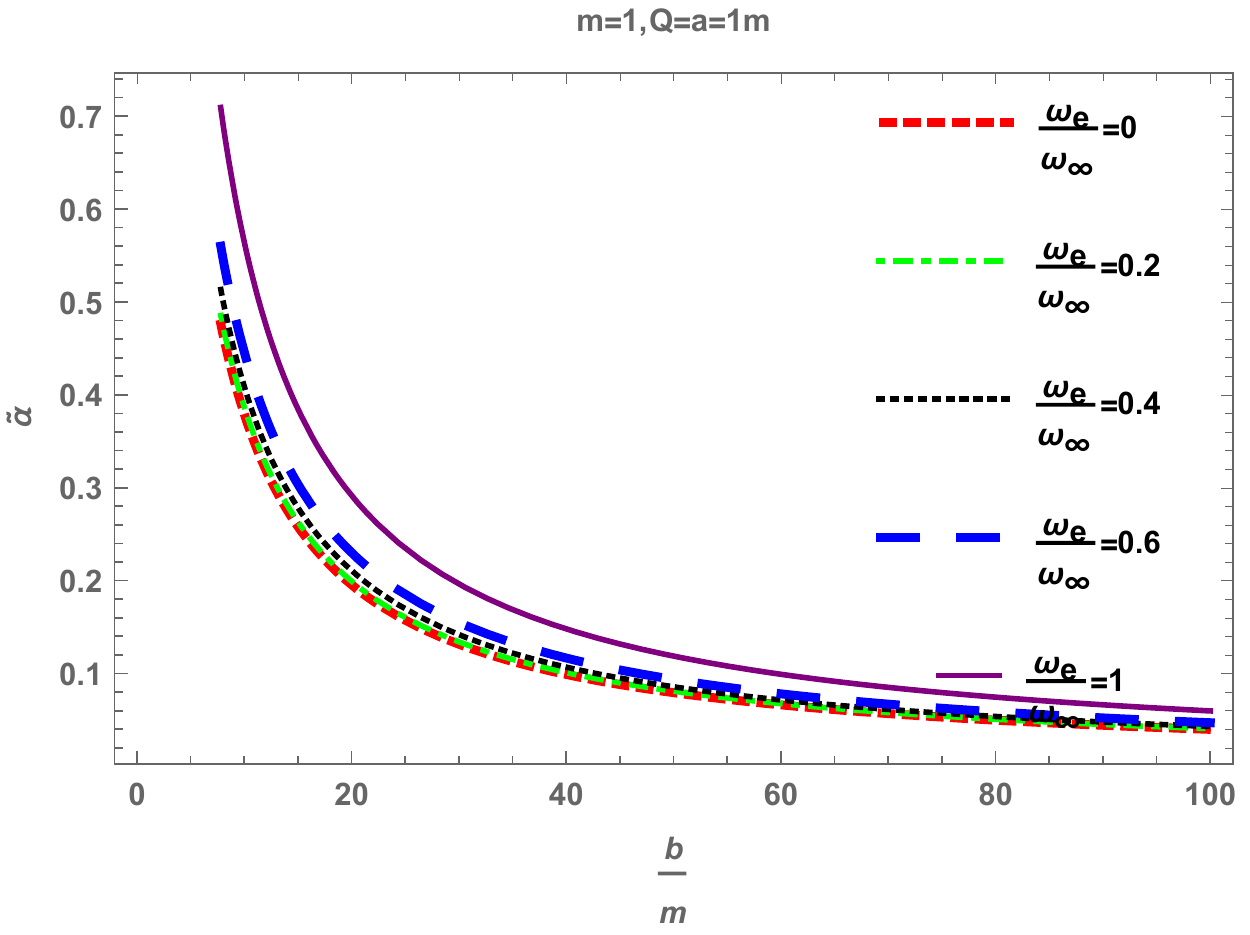}
    \caption{Bending angle's variation $\tilde{\alpha}$ as a function of impact parameter $b$, $Q=a=m$.}
    \label{fig3}
\end{figure}

Figure \ref{fig3} exhibits the behaviour of deflection angle $\tilde{\alpha}$ with respect to impact parameter $b$ for $Q=a=1$ and $0 \leq \frac{\omega_{e}}{\omega_{\infty}} \leq 1$. {We examine that the deflection angle $\tilde{\alpha}$ decreases exponentially and almost approaches {to zero} as the value of impact parameter $b$ goes to infinity. }
In both cases ($Q=a=0.5,1$), angle behavior is stable. It {is observed} that the for $Q=1$ deflection angle $\tilde{\alpha}$ {relative to} the impact parameter $b$ by varying bounce parameter $a$ shows that similar {behavior} as for the $Q=0.5$.
\unskip

\section{Dark Matter's Influence on Deflection Angle} \label{sec5}
This section primarily concerns the calculations of black bounce Reissner--Nordstr\"{o}m BH's deflection angle in DM medium. The dark-atom concept has been offered as a composite of DM, which we study here using light's bending phenomenon. Dark matter possesses electromagnetic interactions due to its frequency-dependent refractive index \cite{Latimer:2013}, and this medium has particular optical characteristics that a traveling photon may detect. The refractive index determines how fast a wave moves through a medium. In this regard, the refractive index for the black bounce Reissner--Nordstr\"{o}m BH is defined as \cite{Latimer:2013}.
\begin{eqnarray} \label{eDM}
n(\omega)=1+\beta A_{0}+ A_{2}\omega^{2}.
\end{eqnarray}

The frequency 
of a photon is represented by $\omega$. Here, it is examined that $\beta=\frac{\rho_{0}}{4m^{2}\omega^{2}}$, where $\rho_{0}$ represents the mass density of dispersed  particles of DM, $A_{0}=-2\varepsilon^{2}e^{2}$ and $A_{2}\geq0$. The optical Gaussian curvature $\mathcal{\tilde {K}}$ of the  black bounce Reissner--Nordstr\"{o}m BH in DM medium up to the leading order terms by using the Equation $\eqref{R44}$ can be calculated as
\begin{eqnarray}
\mathcal{\tilde {K}} &\simeq& \frac{3 Q^2}{r^4 (1+A_{2}\omega^{2}+A_{0} \beta)^2}-\frac{a^2}{r^4 (1+A_{2}\omega^{2}+A_{0}\beta)^2}-\frac{12
Q^2 a^2}{r^6 (1+A_{2}\omega^{2}+A_{0}\beta)^2}\nonumber
\\&-&\frac{2 m}{r^{3} (1+A_{2}\omega^{2}+A_{0}\beta)^2}-\frac{6 Q^2 m}{r^{5}(1+A_{2}\omega^{2}+A_{0}\beta)^2}\nonumber
\\&+&\frac{10a^2 m}{r^5(1+A_{2}\omega^{2}+A_{0}\beta)^2}+\frac{28 Q^{2}a^2 m}{r^7(1+A_{2}\omega^{2}+A_{0}\beta)^2}+\mathcal{O}(m^{2},a^{4},Q^{4}).\label{R13}
\end{eqnarray}

The bending angle $\tilde{\alpha}$ black bounce Reissner--Nordstr\"{o}m BH in DM medium by using Equations $\eqref{R13}$ and $\eqref{R10}$ up to the leading order terms can be computed as
\begin{eqnarray}
\tilde{\alpha} &\simeq& \frac{4m}{b(1+A_{2}\omega^{2}+A_{0} \beta)^2}-\frac{8a^{2} m}{3b^3(1+A_{2}\omega^{2}+A_{0} \beta)^2}+\frac{a^2 \pi}{4b^2 (1+A_{2}\omega^{2}+A_{0} \beta)^2}\nonumber
\\&+&\frac{64a^2 m Q^2}{15b^5 (1+A_{2}\omega^{2}+A_{0} \beta)^2}-\frac{8m Q^2}{3b^3(1+A_{2}\omega^{2}+A_{0} \beta)^2}+\frac{27 a^2 \pi Q^2}{32b^4 (1+A_{2}\omega^{2}+A_{0} \beta)^2}\nonumber
\\&-&\frac{3\pi Q^2}{4b^2(1+A_{2}\omega^{2}+A_{0} \beta)^2}-\frac{16 a^2 m A_{2}\omega^{2}}{3b^3 (1+A_{2}\omega^{2}+A_{0} \beta)^2}+\frac{8 m A_{2}\omega^{2}}{b (1+A_{2}\omega^{2}+A_{0} \beta)^2}\nonumber
\\&+&\frac{a^2 \pi A_{2}\omega^{2}}{2b^2 (1+A_{2}\omega^{2}+A_{0} \beta)^2}+\frac{128 a^2 m Q^2 A_{2}\omega^{2}}{15b^5 (1+A_{2}\omega^{2}+A_{0} \beta)^2}-\frac{16 m Q^2  A_{2}\omega^{2}}{3b^3(1+A_{2}\omega^{2}+A_{0} \beta)^2}\nonumber
\\&+&\frac{27 a^2 \pi Q^2 A_{2}\omega^{2}}{16b^4 (1+A_{2}\omega^{2}+A_{0} \beta)^2}-\frac{3\pi Q^2 A_{2}\omega^{2}}{2b^2 (1+A_{2}\omega^{2}+A_{0} \beta)^2}
\\&+&\mathcal{O}(m^{2},a^{4},Q^{4},A^{2}_{2},\omega^{4}).\label{R14}
\end{eqnarray}

The BH's mass $m$, charge $Q$, bounce parameter $a$, impact parameter $b$, and DM parameters all are the parameters of the measured deflection angle in Equation $\eqref{R14}$. It is to be observed that the photon deflected through the DM around the black bounce Reissner--Nordstr\"{o}m BH has a large bending angle as compared to the vacuum case \cite{Guo:2022hjp}. By eliminating the DM effect, the angle Equation $\eqref{R14}$ reduces to the bending angle in the case of vacuum. By considering $Q \ne 0$ and $a=0$ in Equation $\eqref{R14}$, one can obtain the expression of the Reissner--Nordstr\"{o}m BH's deflection angle.
 We also find that taking charge $Q=0$ and $a=0$ in Equation $\eqref{R14}$ the obtained angle reduces to the Schwarzschild BH's deflection angle in DM medium.

\section{Hawking Radiation} \label{sec6}
In this part, we use a topological technique based on the GBT and Euler characteristic to derive the Hawking temperature of a black bounce Reissner--Nordstr\"{o}m BH. To derive {the Hawking} temperature using the topological approach, one can utilize the Wick rotation \cite{PhysRevD.15.2738} to use the Euclidean geometry of the two-dimensional spacetime without missing any facts from the four-dimensional spacetime. The spherically static symmetric spacetime of black bounce Reissner--Nordstr\"{o}m BH is defined in Equation $\eqref{M1}$.

{Rewriting the four-dimensional metric into the two-dimensional coordinates by using the Wick rotation condition i.e., } $(\theta=\frac{\pi}{2})$ and $(\tau=it)$
\begin{equation}
ds^2=\left(1-\frac{2m}{\sqrt {r^{2}+a^{2}}}+\frac{Q^{2}}{r^{2}+a^{2}}\right){d\tau^2}
+\frac{1}{\left(1-\frac{2m}{\sqrt {r^{2}+a^{2}}}+\frac{Q^{2}}{r^{2}+a^{2}}\right)}dr^2. \label{R16_2}
\end{equation}

The formula to 
compute the Hawking temperature $T_{H}$ of black bounce Reissner--Nordstr\"{o}m BH after using all the values of the physical constants is defined as \cite{Ali:2020}
\begin{equation}
T_{H} = \frac{1}{4\pi \mathcal{X}}\int_{r_{h}}\sqrt{g} \mathcal{R}dr,\label{R17}
\end{equation}
where, $g=1$ is the determinant of Equation \eqref{R16_2} and $r_{h}$ is the event horizon. Using the values of  Ricci scalar $R$, Euler characteristic $\mathcal{X}=1$ and integrating along the event horizon, the  Hawking temperature $T_{H}$ of black bounce Reissner--Nordstr\"{o}m BH is calculated as
\begin{eqnarray}
T_{H}&=\frac{\sqrt{\left(m+\sqrt{m^{2}-Q^{2}}\right)^{2}-a^{2}} \sqrt{m^{2}-Q^{2}}}{2 \pi \left(m+\sqrt{m^{2}-Q^{2}}\right)^{3}}.\label{R18}
\end{eqnarray}

\begin{figure}[h]
\centering    \includegraphics[width=0.75\textwidth]{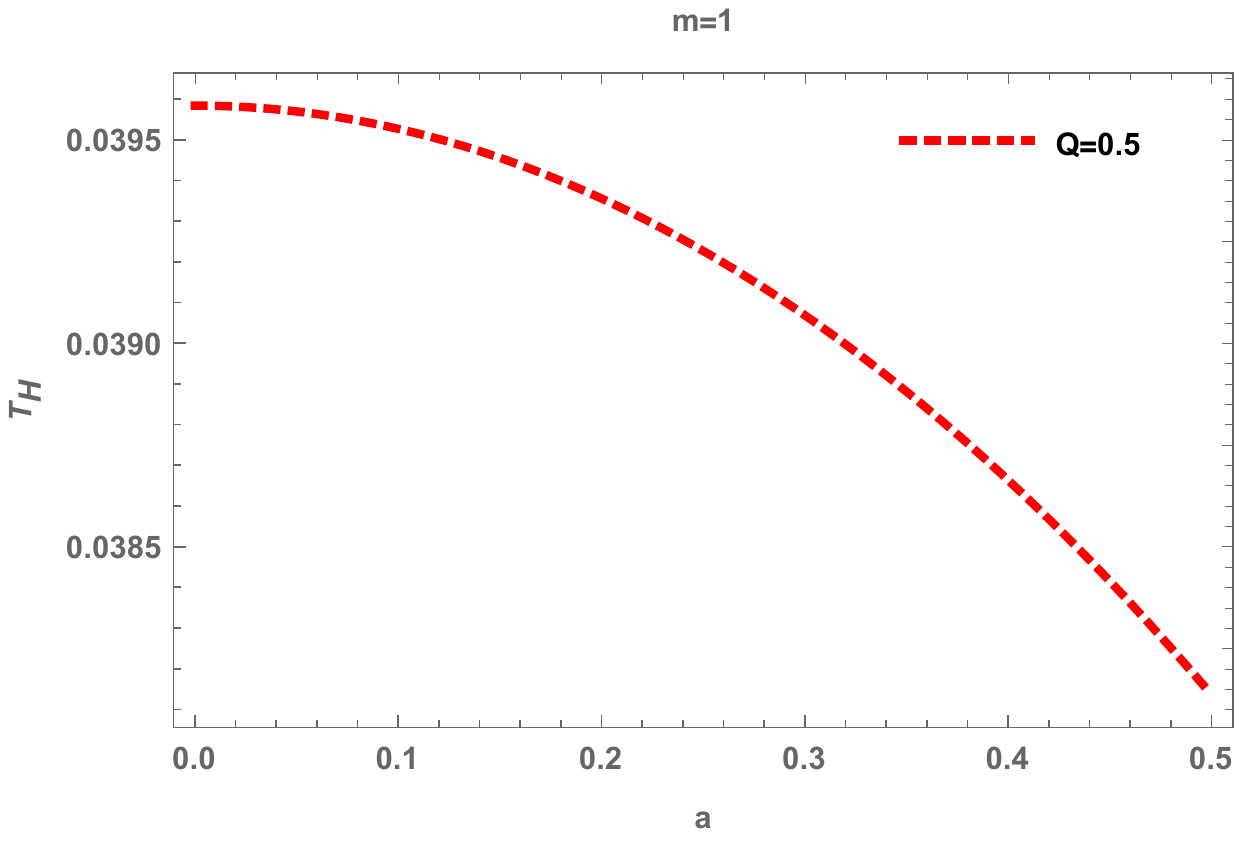}
    \caption{Hawking Temperature $T_{H}$ vs the bounce parameter $a$.}
    \label{fig4n}
\end{figure}

One can observe that the obtained 
expression of the Hawking temperature $T_{H}$ black bounce Reissner--Nordstr\"{o}m BH depends on the mass $m$, charge $Q$ of the BH, and bounce parameter $a$ similarly with \cite{Franzin}. We also notice that the Hawking temperature via standard technique gives the same expression as the topological technique. For the case $Q \ne 0$ and $a=0$, the obtained Hawking temperature in Equation $\eqref{R18}$ reduces to the Hawking temperature of Reissner--Nordstr\"{o}m BH. Moreover, the attained Hawking temperature Equation~$\eqref{R18}$ reduces to the Schwarzschild--Hawking temperature, i.e., $T_{H}=\frac{1}{8m\pi}$ by taking $Q=a=0$. {To observe the {behavior} of Hawking temperature graphically, we plot the graph between Hawking temperature $T_{H}$ and bounce parameter $a$ in Figure \ref{fig4n}. We observe that for $Q=0.5$ the Hawking temperature decreases exponentially.}

\section{Greybody Factor} \label{sec7}
This section mainly examines the greybody factor bound of the black bounce Reissner--Nordstr\"{o}m BH. Much research has been dedicated to estimating the greybody factors. There is, however, a distinct analytic approach for obtaining bounds on the greybody components. The line-element for the  Reissner--Nordstr\"{o}m BH corrected by bounce parameter in a static spherically symmetric spacetime is defined in Equation $\eqref{M1}$
The lower bounds on transmission probability $T$ can be defined as \cite{Visser:1999,BOONSERM20082779,P:2009}.
\begin{eqnarray}
T \geq \sech^{2}\left(\frac{1}{2\omega}\int^{\infty}_{-\infty}\varrho dr_{*}\right),\label{MR4}
\end{eqnarray}
where
\begin{equation}
\varrho = \frac{\sqrt{[g'(r_{*})]^2+[\omega^2-\mathcal{V}(r_{*})-g^2(r_{*})]^2}}{2g(r_{*})}.\nonumber
\end{equation}
where $r_{*}$ represents the tortoise coordinate, and $g$ is a positive function. For the radial {part,} the equation of motion is given as
\begin{equation}
\frac{1}{h(r)^2}\frac{d}{dr}\left[h(r)^2f(r)\frac{du(r)}{dr}\right]+\left[\frac{\omega^{2}}{f(r)}-\frac{l(l+1)}{h(r)^2}\right]{u(r)}=0,
\end{equation}
{where $u(r)$ indicates the  the scalar or vector field oscillating.} Taking $dr_{*}=\frac{1}{f(r)}dr$, while the potential is defined as \cite{Tritos:2013}
\begin{equation}
\mathcal{V}(r)=\frac{l(l+1) f(r)}{h(r)^2}.\label{Rr2}
\end{equation}

The lower bounds 
on the transmission probability $T$ for $g=\omega$ {are} given by
\begin{equation}
T \geq \sech^{2}\left(\frac{1}{2\omega}\int^{\infty}_{r_{h}}\mathcal{V}(r)dr_{*}\right).\label{e25}
\end{equation}

After substituting in 
the values of $\mathcal{V}$ and $dr_{*}$, we obtain the following expression
\begin{equation}
T \geq \sech^{2}\left(\frac{1}{2\omega}\int^{\infty}_{r_{h}}\frac{l(l+1)}{h(r)^2}dr\right).\label{e26}
\end{equation}

The greybody bound  
$T_{b}$ of the black bounce Reissner--Nordstr\"{o}m BH after putting the value of $h(r)^2$  and integrating along $r_{h}$ is calculated as
\begin{equation}
T_{b}=T \geq \sech^{2}\left[\frac{1}{2\omega}\left(\frac{l(l+1)\pi}{2a}-\frac{l(l+1)\arctan \left[\frac{\sqrt{(m+\sqrt{m^{2}-Q^{2}})^{2}-a^{2}}}{a}\right]}{a}\right)\right]
\end{equation}
The bound $T_{b}$ of the black bounce Reissner--Nordstr\"{o}m BH depends upon the mass $m$, charge $Q$, and bounce parameter $a$ of the BH. Guo and Miao \cite{Guo:2022hjp} have also calculated the greybody factor of perturbation fields of the black bounce Reissner--Nordstr\"{o}m BH. It is observed from the graphs that when the potential of the black bounce Reissner--Nordstr\"{o}m BH is higher, then the bound will be lower.

\subsection{Graphical Analysis}
The purpose of this section is to explain the graphical behavior of greybody bound $T_{b}$ and the potential of the black bounce Reissner--Nordstr\"{o}m BH. For this purpose, we take the fixed values of charge $Q$, angular momentum $l=1,2$, and varying bounce parameter $a$.

\begin{figure}[h]
       \includegraphics[width=0.48\textwidth]{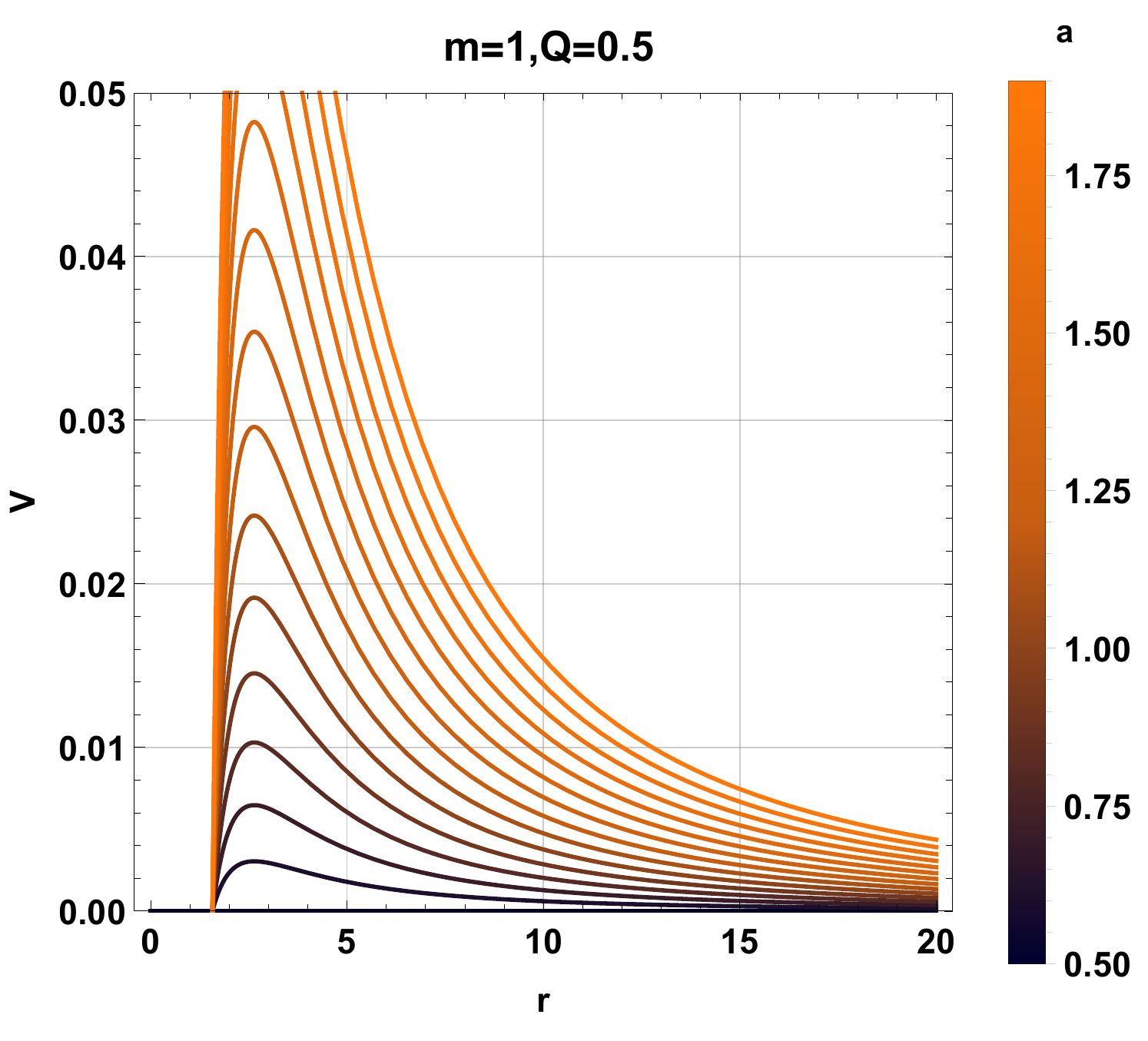}
    \includegraphics[width=0.48\textwidth]{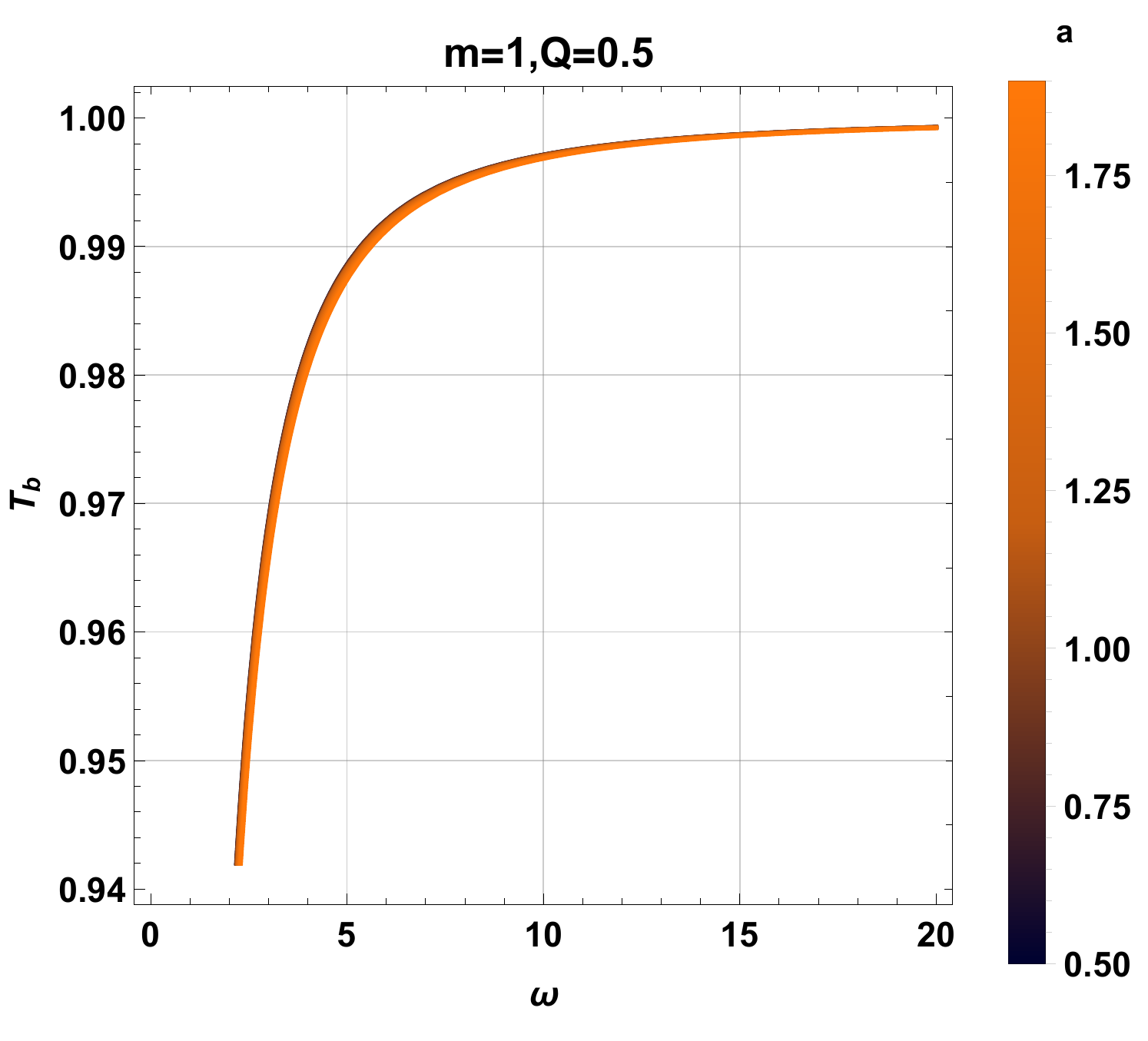}
    \caption{The left panel shows the potential with $l=1$ and corresponding bound $T_{b}$ is shown in right~panel.}
    \label{fig4}
\end{figure}

Figure \ref{fig4} depicts the graphical behavior of the potential $V$ {relative to} the $r$, and greybody factor bound  $T_{b}$ {relative to} the $\omega$. For $0< a < 2$, the potential $\mathcal{V}$ increases, and {attains} its maximum value. However, as the value of bounce parameter $a \rightarrow 0$, the potential exponentially decreases and approaches zero. It is also observed that as the $r \rightarrow 0$, the value of potential is high and attains its maximum value, while for the large values of $r$ the potential starts decreasing from the maximum value {and almost} approaches {zero}. Nevertheless, as the value of $a$ increases, the corresponding bound becomes lower, making it more difficult for the waves to pass through the higher potential. However, the bound $T_{b}$  shows the convergent behavior by converging to $1$.

\begin{figure}[h]
    \includegraphics[width=0.48\textwidth]{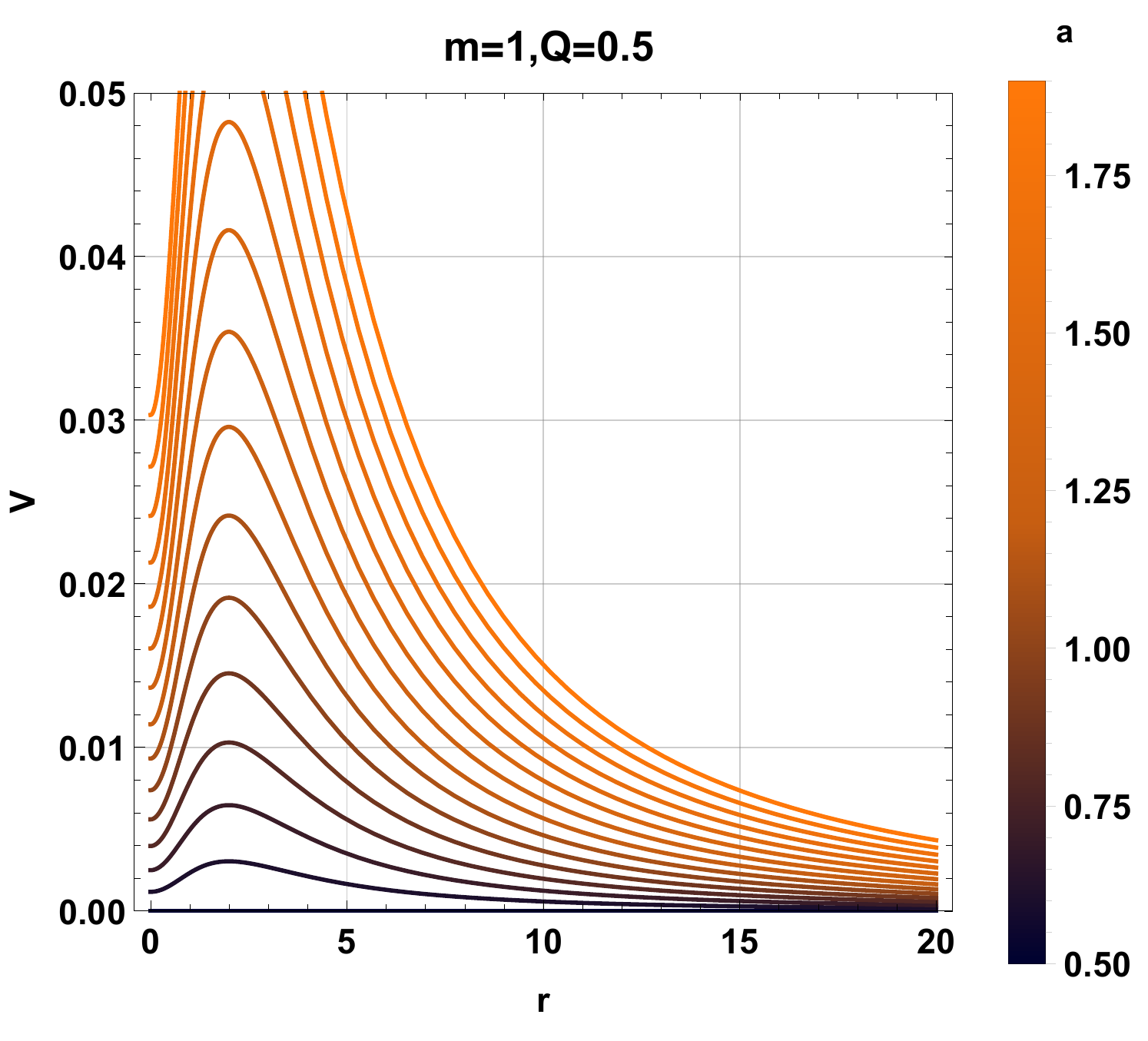}
    \includegraphics[width=0.48\textwidth]{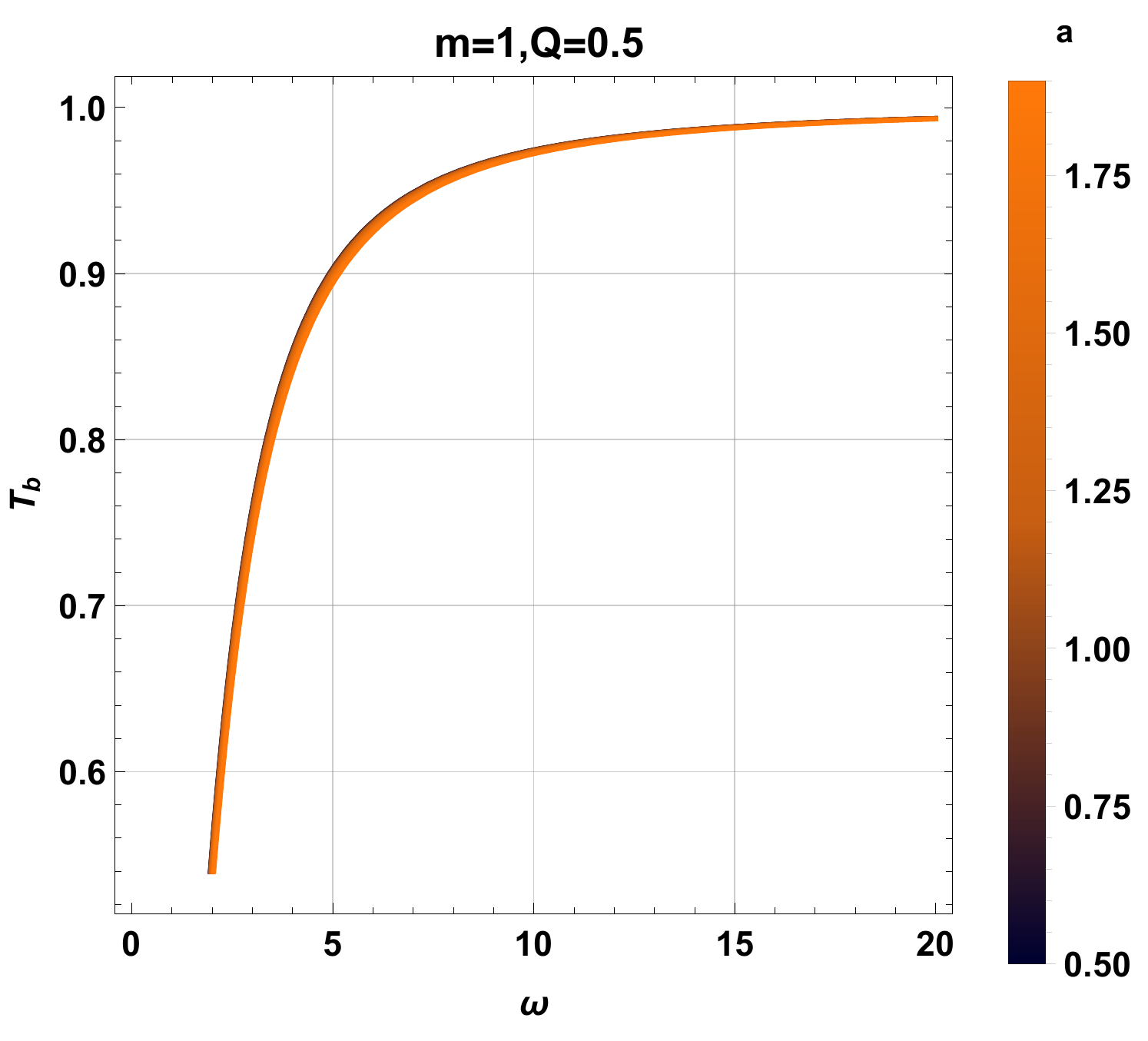}
    \caption{The left panel shows the potential with $l=2$ and corresponding bound $T_{b}$ is shown in right~panel.}
    \label{fig5}
\end{figure}

Figure \ref{fig5} represents the graphical {behavior} of the potential $\mathcal{V}$ with respect to the $r$ and greybody factor bound $T$ with respect to the $\omega$. For $0< a < 2$, the bound $T$ shows {a similar behavior} as for the $l=1$.

\section{Shadow behavior} \label{sec8}
We now turn our attention to the explore the shadow behavior of the black bounce RN black hole. Let us consider the Hamiltonian for light rays where a non-magnetized cold plasma with plasma frequency $\omega_p(r)$ is included \cite{Perlick2015}:
\begin{equation}
    H = \frac{1}{2} \left(g^{ik} p_{i} p_{k} + \omega_p(r)^2 \right) = \frac{1}{2} \left( -\frac{p_{t}^{2}}{A(r)} + \frac{p_{r}^{2}}{B(r)} + \frac{p_{\phi }^{2}}{C(r)} +\omega_p(r)^2\right).
\end{equation}

In the equation above, 
note that {$C(r) = h(r)^2$} due to Equation \eqref{M1}. {Furthermore, we should also note that $A(r) = f(r)$, and $B(r)=A(r)^{-1}$}. Without compromising generality, we can also restrict ourselves along the equatorial plane ($\theta = \pi/2$) due to spherical symmetry and derive the equations of motions (EoS) through the following 
\begin{equation}
    \dot{x}^{i} = \frac{\partial H}{\partial p_{i}}, \quad \quad \dot{p}_{i} = -\frac{\partial H}{\partial x^{i}},
\end{equation}
which reveals two constants of motion:
\begin{equation}
    E = A(r)\frac{dt}{d\lambda}, \quad L = h(r)^2\frac{d\phi}{d\lambda}.
\end{equation}

With the above equation, 
we can define the impact parameter as
\begin{equation}
    b \equiv \frac{L}{E} = \frac{h(r)^2}{A(r)}\frac{d\phi}{dt},
\end{equation}
and the condition that $ds^2 = 0$, gives the rate of change of the $r$-coordinate with respect to the azimuthal angle $\phi$:
\begin{equation}
    \left(\frac{dr}{d\phi}\right)^2 =\frac{h(r)^2}{B(r)}\left(\frac{p(r)^2}{b^2}-1\right),
\end{equation}
where \cite{Perlick2015}
\begin{equation}
    p(r)^2 = \frac{h(r)^2}{A(r)}n(r)^2=\frac{h(r)^2}{A(r)}\left(1-\frac{\omega_e^2}{\omega_\infty^2}A(r) \right)
\end{equation}
since the non-gravitating plasma is assumed to be non-homogenous. With our metric functions, the condition $p'(r) = 0$ allows one to find the photonsphere radius \cite{Perlick2015}, and for our case,
\begin{equation}
    \left(\frac{\omega_{e}^{2}}{\omega_{0}^{2}}A(r)^{2}-A(r)\right)h'(r)^2+h(r)^2A'(r)=0.
\end{equation}

With the inclusion  
of the plasma parameter, finding the analytical expression for $r_\text{ph}$ can be quite lengthy. However, for the case where there is no plasma {(i.e., $n(r)=1$),} we simply found the physical solutions as
\begin{equation} \label{erph}
    r_\text{ph}=\frac{\sqrt{18 M^{2}-8 Q^{2}-4 a^{2}+6 \sqrt{9 M^{4}-8 M^{2} Q^{2}}}}{2}.
\end{equation}

A static observer at a distance $r_\text{obs}$ from the black bounce black hole can obtain the angular radius $\alpha_{\text{sh}}$ of the shadow. {From the black hole's center to $r_\text{obs}$, simple geometry shows that $\Delta x = \sqrt{B(r)} dr$ and $\Delta y = h(r) d\phi$} \cite{Perlick2015}:
\begin{equation}
    \tan(\alpha_{\text{sh}}) = \lim_{\Delta x \to 0}\frac{\Delta y}{\Delta x} = h(r)\left(\frac{1}{B(r)}\right)^{1/2} \frac{d\phi}{dr} \bigg|_{r=r_\text{obs}},
\end{equation}
which can be simplified in terms of the critical impact parameter as
\begin{equation}
    \sin^{2}(\alpha_\text{sh}) = \frac{b_\text{crit}^{2}}{p(r_\text{obs})^{2}}.
\end{equation}

Here, the $b_\text{crit}$ can be obtained using the orbit equation \cite{Pantig:2022ely}: 
\begin{equation}
    b_\text{crit}^2 = \frac {p (r_\text{ph})  \left( 2h (r_\text{ph})^2 B (r_\text{ph}) p'(r_\text{ph})-h (r_\text{ph})^2 B'(r_\text{ph}) p (r_\text{ph}) +B (r_\text{ph}) h' (r_\text{ph})^2 p (r_\text{ph})  \right) }{B (r_\text{ph}) h'(r_\text{ph})^2 - h (r_\text{ph})^2 B'(r_\text{ph})}
\end{equation}

The critical impact parameter's analytical expression with $n(r)$ is somewhat complicated, but for the case {$n(r) = 1$}, we find
\begin{equation}
    b^2_\text{crit} = \frac{2h(r_\text{ph})^3}{h(r_\text{ph})-m}.
\end{equation}

Finally, it can be easily shown that, in terms of $r_\text{obs}$ and $r_\text{ph}$, we obtain the shadow radius as {($n(r) = 1$)}
\begin{equation} \label{eRsh}
    R_\text{sh}=\left[\frac{2h(r_\text{ph})^3(h(r_\text{obs})^2-2mh(r_\text{obs})+Q^2)}{h(r_\text{obs})^2(h(r_\text{ph})-m)}\right]^{1/2}.
\end{equation}

\begin{figure}[h]
\centering       \includegraphics[width=0.75\textwidth]{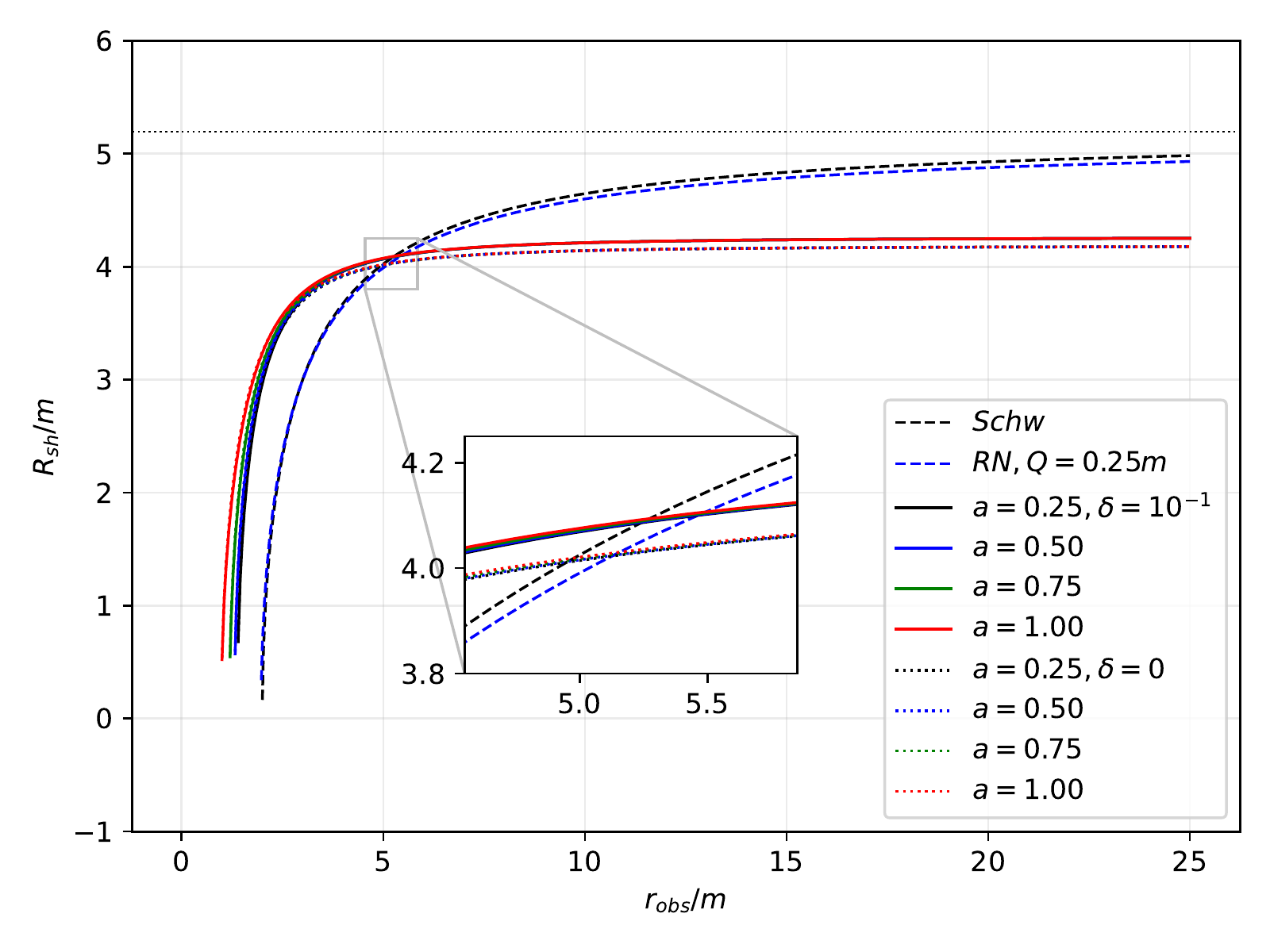}
    \caption{behavior of the shadow radius is due to a static observer with varying location from the black bounce RN BH. Here, we used $Q=0.25m$.}
    \label{sharad}
\end{figure}

Next, we plotted Equation \eqref{eRsh}, which is indicated by the dotted lines in Figure \ref{sharad}. We also included in the plot the case where the black hole bounce is surrounded by plasma (solid lines). For immediate comparison, we also plotted the Schwarzschild and RN cases.
Furthermore, while it is understood that the shadow cast by a non-spinning black hole is a circle, we can see in the plot the behavior of the shadow radius. First, without the bounce parameter, the RN case with $Q = 0.25m$ decreases the shadow radius while following the general trend of the Schwarzschild case. For the Schwarzschild case, {$R_\text{sh} = 3\sqrt{3}m$} as $r_\text{obs}$ becomes larger, but at $r_\text{obs} = 25m$, we observe that it is still lower than this value of $R_\text{sh}$. The effect of the bounce parameter is to produce lower values of $R_\text{sh}$ and {make} its rate of change with respect to $r \to 0$ at lower values of $r_\text{obs}$. It means that the observer does not need to go so far away to observe a constant shadow radius. Furthermore, the bounce parameter allows the formation of the shadow near the event horizon. With the plasma medium $\delta = 10^{-1}$, we observe that $r_\text{obs}$ follows the general trend for $\delta = 0$, but increases the shadow slightly. Such an increase depends on the value of the plasma parameter.

For completion, let us analyze the effect of the dark matter refractive index $n(\omega)$ in Equation \eqref{eDM} instead of the plasma. The photonsphere can be found via
\begin{equation}
    n(\omega)^{2}[h'(r)A(r)-A'(r)h(r)]=0,
\end{equation}
which reveals that the photonsphere radius is independent of the dark matter parameter. It is easy to see that through the orbit equation, one can verify that the critical impact parameter in this case is
\begin{equation}
    b^2_\text{crit} = \frac{2 n(\omega)^2 h(r_\text{ph})^3}{h(r_\text{ph})-m}.
\end{equation}

Then, the shadow is given by
\begin{equation}
    R_\text{sh}=n(\omega)h(r_\text{ph})\left[\frac{2 h(r_\text{ph})(h(r_\text{obs})^2-2mh(r_\text{obs})+Q^2)}{h(r_\text{obs})^2(h(r_\text{ph})-m)}\right]^{1/2},
\end{equation}
where the shadow radius is increased by a factor of $n(\omega)$.

\section{Conclusion} \label{conclu}
In this work, we have discussed the Reissner--Nordstr\"{o}m BH corrected by bounce parameter and its properties, i.e., the curvature singularities are absent from the black bounce family on a global scale and satisfy all observable weak field tests. In {the case} of plasma and DM mediums, the attained bending angle Equation $\eqref{R11}$ depended on the mass $m$, charge $Q$ of the BH, bounce parameter $a$, impact parameter, and medium's parameters. It is noted that in bending angle Equation $\eqref{R11}$ the terms without the bounce parameter $a$ and which {contain} charge are due to the charged nature of the BH and the remaining terms are due to the corrections with the bounce parameter $a$. It is also found that the effect of the plasma increases the deflection angle. The bending angle is inversely proportional to the photon frequency, so the bending angle increases by lowering the photon frequency and assuming the electron frequency is fixed.
It is to be observed that black bounce Reissner--Nordstr\"{o}m BH's bending angle increases due to the DM medium effect compared to the vacuum case.

In above both mediums, we have examined that in the absence of a bounce parameter, one can obtain the bending angle of Reissner--Nordstr\"{o}m BH and the bending angle of Schwarzschild's BH  by neglecting the charge and bounce parameter of the BH. Moreover, one can attain the bending angle of a non-plasma medium by taking $\omega_{e}=0$ in a plasma medium. It is noticed that also that, by ignoring the DM effect in the angle of deflection in Equation $\eqref{R11}$, the angle reduces to the angle of a non-plasma medium. We also observed that the obtained bending angle in both mediums is directly proportional to the mass $m$, charge $Q$, bounce parameter $a$, and inversely proportional to the impact parameter $b$.

Following that, we have analyzed the graphical behavior of the bending angle $\tilde{\alpha}$ with respect to the $b$ for fixed values of mass $m$, charge $Q$, $\frac{\omega_{e}}{\omega_{\infty}}=0.1$ and varying the value of bounce parameter $a$. {We have examined that for at the small values of impact parameter b, the value of the bending angle  $\tilde{\alpha}$ is maximum and as the value of $b$ increases, the bending angle  $\tilde{\alpha}$ exponentially decreases and approaches to zero.} Moreover, we have investigated the deflection angle $\tilde{\alpha}$ with respect to the impact parameter $b$ by fixing $m$, $Q$, taking $a=0.5,1$ and varying the plasma factor. {For $Q=a=0.5,1$, we have studied that the deflection angle $\tilde{\alpha}$ decreases exponentially and almost approaches {zero} as the value of impact
parameter $b$ goes to infinity.  In all {the above} cases, graphically we have observed that the deflection angle  $\tilde{\alpha}$ shows the inverse {relationship} with the impact parameter $b$ and also the {behavior} of angle is physically stable.}

Furthermore, we have computed the Hawking temperature using a topological method involving two invariants, namely the two-dimensional Euler characteristic and the GBT. We have examined that the obtained expression of the Hawking temperature $T_{H}$ in Equation~$\eqref{R18}$  black bounce Reissner--Nordstr\"{o}m BH depends on the mass $m$, charge $Q$ of the BH and bounce parameter $a$. We also noticed that the Hawking temperature expression is similar to the topological technique. It is to be mentioned here that for the case $Q \ne 0$ and $a=0$, the Hawking temperature Equation $\eqref{R18}$ reduces to the Hawking temperature of Reissner--Nordstr\"{o}m BH, and for $Q=a=0$, the acquired Hawking temperature converts to the Schwarzschild  Hawking temperature, i.e., $T_{H}=\frac{1}{8m\pi}$. {Furthermore, we graphically investigated that the Hawking {temperature decreases} exponentially.}

We have also calculated the greybody bound $T_{b}$ and examined that the bound $T_{b}$ of the black bounce Reissner--Nordstr\"{o}m BH depends on the mass $m$, charge $Q$ of the BH, and bounce parameter $a$. Moreover, we have observed that the potential $\mathcal{V} $ increases and attains its maximum value for $l=1,2$. As the value of $a$ increases, the potential exponentially decreases and approaches zero. It is to be mentioned here that when $r \rightarrow 0$, one can attain the high value of potential, and for the large value of $r$ the potential approaches zero. It is observed that the corresponding bound becomes lower as the value of $a$ increases. Furthermore, it is examined that the greybody factor's bound exhibits the convergent behavior and converges to $1$. We also observed that for large values of $a$ and small $r$, the potential is higher, making it difficult for the waves to pass through that potential.

Finally, we also explored the effect of the bounce parameter on the behavior of the shadow radius and when it is surrounded by plasma. First, the effect of the bounce parameter is to allow shadow formation closer to the black hole shadow and at a larger radius than the Schwarzschild or RN cases. Here, the rate at which the shadow increases is also larger. Moreover, we observe that the bounce parameter quickly makes the shadow radius rate of change tend to zero even at low values of $r_\text{obs}$. Finally, the effect of the plasma is just to increase the shadow radius of the black hole affected by the bounce parameter. These parameters can indeed change the shadow radius, which sophisticated astronomical devices can detect.

\begin{acknowledgements}
 A. {\"O}. and R. P. would like to acknowledge networking support by the COST Action CA18108 - Quantum gravity phenomenology in the multi-messenger approach (QG-MM).
\end{acknowledgements}

\bibliography{ref}

\end{document}